\documentclass[onecolumn]{svjour2}

\smartqed

\usepackage{graphicx}

\journalname{J Stat Phys}

\begin{document}

\title{Charge and Current Sum Rules in Quantum Media Coupled to Radiation II}

\titlerunning{Charge and Current Sum Rules in Quantum Media II}

\author{Ladislav \v{S}amaj \and Bernard Jancovici}

\institute{L. \v{S}amaj \at \\ Institute of Physics,
                Slovak Academy of Sciences, D\'ubravsk\'a cesta 9,
                845 11 Bratislava, Slovakia \\
                \email{Ladislav.Samaj@savba.sk} \\
                \and B. Jancovici \at \\
                Laboratoire de Physique Th\'eorique, Universit\'e
                de Paris-Sud, 91405 Orsay Cedex, France (Unit\'e Mixte
                de Recherche No 8627 - CNRS)  \\
                \email{Bernard.Jancovici@th.u-psud.fr}}
\date{Received:  / Accepted: }

\maketitle

\begin{abstract}
This paper is a continuation of the previous study [\v{S}amaj, L.:
J. Stat. Phys. {\bf 137}, 1-17 (2009)], where a sequence of sum rules 
for the equilibrium charge and current density correlation functions
in an infinite (bulk) quantum media coupled to the radiation
was derived by using Rytov's fluctuational electrodynamics.
Here, we extend the previous results to inhomogeneous situations,
in particular to the three-dimensional interface geometry of 
two joint semi-infinite media.
The sum rules derived for the charge-charge density correlations represent
a generalization of the previous ones, related to the interface dipole moment 
and to the long-ranged tail of the surface charge density correlation
function along the interface of a conductor in contact with an inert 
(not fluctuating) dielectric wall, to two fluctuating semi-infinite media 
of any kind.
The charge-current and current-current sum rules obtained here are, 
to our knowledge, new.
The current-current sum rules indicate a breaking of the directional 
invariance of the diagonal current-current correlations by the interface. 
The sum rules are expressed explicitly in the classical high-temperature 
limit (the static case) and for the jellium model (the time-dependent case).

\keywords{Sum rules \and inhomogeneous systems \and fluctuations 
\and radiation \and classical limit \and jellium}

\end{abstract}

\renewcommand{\theequation}{1.\arabic{equation}}
\setcounter{equation}{0}

\section{Introduction} \label{Sect.1}
The models studied in this paper are composed of spinless charged particles,
classical or quantum, which are non-relativistic, i.e. they behave
according to Schr\"odinger and not Dirac.
On the other hand, the interaction of charged particles via the radiated
electromagnetic (EM) field can be considered either non-relativistic
(nonretarded) or relativistic (retarded).
In the nonretarded regime, magnetic forces are ignored by taking
the speed of light $c\to\infty$, so that the particles interact only
via instantaneous Coulomb potentials.
In the retarded regime, $c$ is assumed finite and the particles are
fully coupled to both electric (longitudinal) and magnetic (transverse)
parts of the radiated field.

One of the tasks in the equilibrium statistical mechanics of charged
systems is to determine how fluctuations of microscopic quantities
like charge and current densities, induced electric and magnetic fields,
etc., around their mean values are correlated in time and space.
A special attention is devoted to the asymptotic large-distance
behavior of the correlation functions and to the sum rules, which
fix the values of certain moments of the correlation functions.  

Two complementary types of approaches exist in the theory of charged systems.
The microscopic approaches, based on the explicit solution of models 
defined by their microscopic Hamiltonians, are usually restricted to 
the nonretarded regime.
A series of sum rules for the charge and current correlation functions 
has been obtained for infinite (bulk), semi-infinite and fully finite
geometries (see review \cite{Martin88}). 
The quantum sum rules are available only for the jellium model of conductors
(sometimes called the one-component plasma), i.e. the system of identically 
charged pointlike particles immersed in a neutralizing homogeneous background,
in which there is no viscous damping of the long-wavelength plasma 
oscillations. 
The macroscopic approaches are based on the assumption of validity of
macroscopic electrodynamics. 
Being essentially of mean-field type, they are expected to provide 
reliable results only for the leading terms in the asymptotic 
long-wavelength behavior of correlations. 
In general, these approaches are able to predict basic features 
of physical systems also in the retarded regime.
A macroscopic theory of equilibrium thermal fluctuations of the EM field
in quantum media, conductors and dielectrics, was proposed by Rytov 
\cite{Rytov53,Levin67,LP}.

In a recent work \cite{Samaj09}, a sequence of static or time-dependent
sum rules, known or new, was obtained for the bulk charge and current
density correlation functions in quantum media fully coupled to the radiation
by using Rytov's fluctuational electrodynamics.
A technique was developed to extract the classical and purely
quantum-mechanical parts of these sum rules.
The sum rules were critically tested on the jellium model.
A comparison was made with microscopic approaches to systems of particles
interacting through Coulomb forces only \cite{Martin85,John93};
in contrast to microscopic results, the current-current density correlation
function was found to be integrable in space, in both classical and
quantum cases. 
 
\begin{figure*} \label{Fig.1}
\includegraphics[width=0.60\textwidth,clip]{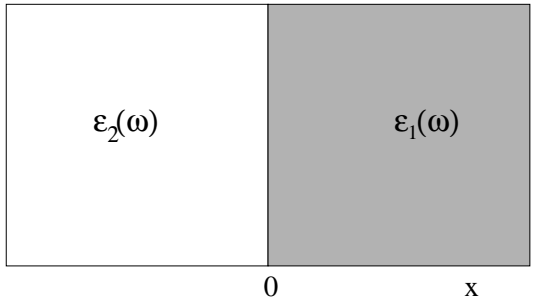}
\caption{Two semi-infinite media characterized by dielectric functions 
$\epsilon_1(\omega)$ and $\epsilon_2(\omega)$}
\end{figure*}

This paper is a continuation of the previous study \cite{Samaj09}. 
It aims at generalizing the previous sum rules to inhomogeneous situations,
in particular to the interface geometry of two semi-infinite media with 
different dielectric functions pictured in Fig. 1.
It should be emphasized that this is not exactly the configuration 
considered in some previous studies.
The standard configuration was a conductor in contact with an ``inert'' 
(not fluctuating) wall of the static dielectric constant $\epsilon_W$.
The presence of a dielectric wall is reflected itself only via the 
introduction of charge images; the microscopic quantities inside 
the inert wall do not fluctuate, they are simply fixed to their mean values.
Such a mathematical model can provide a deformed description of real
materials and, as is shown in this paper, it really does.
The only exception from the described inert-wall systems
is represented by the specific (two-dimensional) two-densities 
jellium, i.e. the interface model of two joint semi-infinite 
jelliums with different mean particle densities, treated in 
\cite{Blum84,Jancovici84,Alastuey84}.
It stands to reason that in the case of the vacuum $(\epsilon_W=1)$ 
plain hard wall, there is no charge which could fluctuate and
the inert-wall model is therefore adequate.

To our knowledge, the sum rules for a (fluctuating) conductor medium 
in contact with a dielectric (inert) wall obtained up to now were 
restricted to the charge-charge density correlation functions.
The inhomogeneous charge-charge sum rules are either of dipole type or
they are related to the long-ranged decay of the surface charge correlation
function along the interface.

The classical dipole sum rule for the static charge-charge density correlations 
follows directly from the Carnie and Chan generalization to nonuniform fluids 
of the second-moment Stillinger-Lovett condition \cite{Carnie81,Carnie83}.
The time-dependent classical dipole sum rule was derived in \cite{Lebowitz85}. 
A time-dependent generalization of the Carnie-Chan rule to the quantum
(nonretarded) jellium and the consequent derivation of the quantum dipole 
sum rule for the time-dependent charge-charge density correlations were 
accomplished in Ref. \cite{Jancovici85b}.

The bulk charge correlation functions exhibit a short-ranged, usually
exponential, decay in classical conductors due to the screening. 
On the other hand, for a semi-infinite conductor in contact with a vacuum 
or (inert) dielectric wall, the correlation functions of the surface charge 
density on the conductor decay as the inverse cube of the distance at 
asymptotically large distances \cite{Usenko79}.
In the classical static case, this long-range phenomenon has been obtained
microscopically \cite{Jancovici82a,Jancovici82b} as well as by using
simple macroscopic argument based on the electrostatic method of
images \cite{Jancovici95}; the prefactor to the asymptotic decay was found
to be universal, i.e. independent of the composition of the Coulomb fluid.
In the quantum case of the specific jellium model, ignoring retardation 
effects, a nonuniversal prefactor to the asymptotic decay was obtained, 
for both static \cite{Jancovici85a} and time-dependent 
\cite{Jancovici85b,Jancovici85a} correlation functions.
Recently \cite{Samaj08,Jancovici09a}, by using the inhomogeneous version
of Rytov's fluctuational theory, we have extended the quantum analysis of
the jellium to the retarded case.
We got a surprising result: for both static and time-dependent surface
charge correlation functions, the inclusion of retardation effects
causes the quantum prefactor to take its universal static classical form,
for any temperature.
The restoration of the classical prefactor by retardation effects
was observed subsequently for arbitrary (conductor, dielectric, vacuum) 
configurations of two semi-infinite quantum media \cite{Jancovici09b}.   

In this paper, we apply the inhomogeneous version of the Rytov fluctuational
theory to extend the bulk sum rules, derived in \cite{Samaj09}, 
to the geometry of two joint semi-infinite media with distinct dielectric 
functions in Fig. 1.
The sum rules derived for the charge-charge density correlations represent
a generalization of the previous (dipole moment and surface charge) ones, 
valid for a conductor system in contact with an inert dielectric wall, 
to two fluctuating semi-infinite media of any kind.
The fundamental differences between the results for the inert and fluctuating 
wall descriptions are pointed out.
The charge-current and current-current sum rules obtained here are, 
to our knowledge, new.
The current-current sum rules indicate a breaking of the directional 
invariance of the diagonal current-current correlations by the interface. 
The sum rules are expressed explicitly in the classical high-temperature 
limit (the static case) and for the jellium model (the time-dependent case).

The paper is organized as follows.
In Sec. 2, we review the inhomogeneous Rytov theory of EM field fluctuations
and write down basic expressions for the charge and current densities;
explicit results for the elements of the retarded Green function tensor
for the two semi-infinite media configuration in Fig.1 are presented 
in Appendix.
Dipole sum rules for the charge-charge density correlation functions
are derived in Sect. 3.
The sum rules related to the long-ranged tail of the surface charge density 
correlation function along the interface between two media are discussed
in Sect. 4 which is divided into three parts.
In the first part 4.1, we generalize the classical static analysis
of a medium in vacuum \cite{Jancovici09b} to arbitrary media configurations.
Part 4.2 concerns the derivation of a classical static relation between 
the dipole moment and the large distance asymptotic of 
the surface charge density.
Part 4.3 is a brief recapitulation of the quantum case, in both retarded
and nonretarded regimes.
The sum rules for the charge-current and current-current density correlation
functions are the subject of Sects. 5 and 6, respectively.
Section 7 is the Conclusion.

\renewcommand{\theequation}{2.\arabic{equation}}
\setcounter{equation}{0}

\section{Fluctuational Formalism} \label{Sect.2}
We consider the (3+1)-dimensional space of points, defined by Euclidean 
vectors ${\bf r}=(x,y,z)$ and time $t$.
We shall deal with semi-infinite geometries, inhomogeneous say along the first 
coordinate $x$; it is useful to denote the remaining two coordinates 
normal to $x$ as ${\bf R}=(y,z)$.
The model consists in two distinct semi-infinite media (conductors, 
dielectrics or vacuum) with the frequency-dependent dielectric functions
$\epsilon_1(\omega)$ and $\epsilon_2(\omega)$ which are
localized in the complementary half spaces 
$\Lambda_1=\{ {\bf r}=(x>0,{\bf R}) \}$ and
$\Lambda_2=\{ {\bf r}=(x<0,{\bf R}) \}$,
respectively, so that the interface between the media
is localized at $x=0$ (see Fig. 1).
We shall assume that the media have no magnetic structure,
i.e. they are not magnetoactive, and the magnetic
permeabilities $\mu_1=\mu_2=1$.
The two-point functions studied in this paper will be translationally
invariant in time and in the vector space ${\bf R}$ perpendicular
to the $x$ axis, and so we shall use the (partial) Fourier representation
\begin{eqnarray}
f(t,{\bf r};t',{\bf r}') & \equiv & f(t-t',{\bf R}-{\bf R}';x,x') 
\nonumber \\ & = & 
\int_{-\infty}^{\infty} \frac{{\rm d}\omega}{2\pi}
\int_{R^2} \frac{{\rm d}{\bf q}}{(2\pi)^2} 
{\rm e}^{-{\rm i}\omega(t-t')+{\rm i}{\bf q}\cdot ({\bf R}-{\bf R}')}
f(\omega,{\bf q};x,x') , \label{2.1}
\end{eqnarray}  
where $\omega$ denotes the frequency and ${\bf q}=(q_y,q_z)$ is the
two-dimensional wave vector.

The induced EM fields inside a material medium are random variables which
fluctuate in time and space due to the random motion of charged particles. 
In the long-wavelength scale, the EM fluctuations are described by 
the Rytov theory \cite{Rytov53,Levin67,LP}.
This theory is usually formulated in the Weyl gauge with the
scalar potential $\phi(t,{\bf r})=0$, so that the (classical) vector 
potential ${\bf A}(t,{\bf r})$ with components $A_j(t,{\bf r})$ $(j=x,y,z)$
determines the microscopic electric and magnetic fields as follows
\begin{equation} \label{2.2}
{\bf E} = - \frac{1}{c} \frac{\partial {\bf A}}{\partial t} , \qquad
{\bf B} = {\rm curl} {\bf A} 
\end{equation}
(we use Gaussian units), where $c$ is the speed of light.
In the context of the quantized EM field theory, the crucial role is played
by the retarded photon Green function tensor $\tens{D}$ defined by  
\begin{equation} \label{2.3}
{\rm i} D_{jk}(t-t';{\bf r},{\bf r}') = \left\{
\begin{array}{lr} 
\langle A_j(t,{\bf r}) A_k(t',{\bf r}') - A_k(t',{\bf r}') A_j(t,{\bf r}) 
\rangle , & t\ge t' ,\cr & \cr 0 , & t<t' ,
\end{array} \right. 
\end{equation}
where $A_j(t,{\bf r})$ denotes the vector-potential operator
in the Heisenberg picture and the angular brackets represent
the equilibrium averaging at temperature $T$, or the inverse
temperature $\beta=1/(k_{\rm B}T)$.
For non-magnetoactive media, the Green function tensor possesses
the symmetry
\begin{equation} \label{2.4}
D_{jk}(t-t';{\bf r},{\bf r}') = D_{kj}(t-t';{\bf r}',{\bf r}) .
\end{equation}
In the Fourier space, the symmetry is expressible as
\begin{equation} \label{2.5}
D_{jk}(\omega,{\bf q};x,x') = D_{kj}(\omega,-{\bf q};x',x) .
\end{equation}

The validity of macroscopic Maxwell's equations for the mean values of 
the EM fields implies, in the frequency Fourier space, 
a set of differential equations of dyadic type fulfilled by 
the Green function tensor:
\begin{equation} \label{2.6}
\sum_l \left[ \frac{\partial^2}{\partial x_j \partial x_l}
- \delta_{jl} \Delta - \delta_{jl} \frac{\omega^2}{c^2} 
\epsilon(\omega,{\bf r}) \right] D_{lk}(\omega;{\bf r},{\bf r}') 
= - 4 \pi \hbar \delta_{jk} \delta({\bf r}-{\bf r}') .
\end{equation} 
Here, in order to simplify the notation, the vector ${\bf r}=(x,y,z)$
is represented as $(x_1,x_2,x_3)$.
The {\em source} point ${\bf r}'$ and the index $k$ only act as 
some fixed parameters, the boundary conditions are with respect
to the {\em field} point ${\bf r}$.
There is an obvious boundary condition of regularity
$D_{jk}(\omega;{\bf r},{\bf r}')\to 0$ for asymptotically large distances
$\vert {\bf r}-{\bf r}'\vert\to\infty$.
At an interface between two different media, the boundary conditions
correspond to the macroscopic requirement that the tangential components
of the fields ${\bf E}$ and ${\bf H}={\bf B}$, considered in the gauge
(\ref{2.2}), be continuous.
The Green function tensor for the geometry pictured in Fig. 1 was obtained 
in a number of papers, see e.g. the method using vector wave functions 
\cite{Li94} or the Weyl expansion method \cite{Tomas95,Dung98}.
The results are usually written in a complicated way, by using
the dyadic notation for the tensors.
In order to enable the reader to reproduce easily calculations performed in 
this work, in the Appendix we present explicitly the Fourier transforms 
(\ref{2.1}) of the tensor elements $D_{jk}(\omega,{\bf q};x,x')$ 
for two possible cases: the points ${\bf r}$ and ${\bf r}'$ are 
in the same half space or they are in different half spaces.

Applying the fluctuation-dissipation theorem and assuming 
the symmetry (\ref{2.5}), the fluctuations of the vector potential 
are described by the formula   
\begin{equation} \label{2.7}
\langle A_j(\omega,{\bf q};x) A_k(-\omega,-{\bf q};x') \rangle^s
= - \coth\left( \frac{\beta\hbar\omega}{2} \right)
{\rm Im}\, D_{jk}(\omega,{\bf q};x,x') ,
\end{equation}
where Im means the imaginary part and
$\langle A_j(\omega,{\bf q};x) A_k(-\omega,-{\bf q};x') \rangle^s$
is the Fourier transform of the symmet\-rized correlation function
\begin{equation} \label{2.8}
\langle A_j(t,{\bf r}) A_k(t',{\bf r}') \rangle^{s} \equiv
\frac{1}{2} \langle A_j(t,{\bf r}) A_k(t',{\bf r}')
+ A_k(t',{\bf r}') A_j(t,{\bf r}) \rangle^{\rm T} .
\end{equation}
Here, $\langle \cdots \rangle^{\rm T}$ represents a truncated
equilibrium average, $\langle AB \rangle^{\rm T} = \langle AB \rangle
- \langle A\rangle \langle B \rangle$.

The relation (\ref{2.7}) enables us to calculate the symmetrized 
two-point correlation function of arbitrary statistical quantities.
Let a scalar quantity $u$ be expressible in terms of the components of 
the vector potential, in the classical format and in the gauge (\ref{2.2}), as
$u(t,{\bf r})= \sum_j {\bf U}_j A_j(t,{\bf r})$, where $\{ {\bf U}_j\}$
$(j=x,y,z)$ are operators acting on time and space variables.
Within the spectral representation with a single frequency $\omega$
and two-dimensional vector ${\bf q}$, $f(t,{\bf r}) = 
{\rm e}^{-{\rm i}\omega t + {\rm i}{\bf q}\cdot {\bf R}} f(\omega,{\bf q};x)$,
this relation takes a partially algebraic form $u(\omega,{\bf q};x) 
= \sum_j {\bf U}_j(\omega,{\bf q};x) A_j(\omega,{\bf q};x)$
(${\bf U}_j$ can still act as operators on the $x$ coordinate).
It follows from the definition (\ref{2.1}) that the Fourier transform
of the symmetrized two-point correlation function of statistical quantities
$u$ and $v$, $\langle u(t,{\bf r}) v(t',{\bf r}') \rangle^s$,
is then determined by
\begin{eqnarray} 
\langle u(x) v(x') \rangle^s_{\omega,{\bf q}} & \equiv & 
\langle u(\omega,{\bf q},x) v(-\omega,-{\bf q},x') \rangle^s \nonumber \\
& = & \sum_{jk} {\bf U}_j(\omega,{\bf q};x) {\bf V}_k(-\omega,-{\bf q};x') 
\langle A_j(x) A_k(x') \rangle^s_{\omega,{\bf q}} , 
\label{2.9}
\end{eqnarray}
where $\langle A_j(x) A_k(x') \rangle^s_{\omega,{\bf q}} \equiv
\langle A_j(\omega,{\bf q};x) A_k(-\omega,-{\bf q};x') \rangle^s$
is given by (\ref{2.7}).
The statistical quantities of our interest are the volume charge density 
$\rho$ and the electric current density ${\bf j}$.
The charge density, defined by $4\pi\rho(t,{\bf r})={\rm div}\, {\bf E}$,
is expressible in terms of the vector potential components as follows
\begin{equation} \label{2.10}
\rho(\omega,{\bf q};x) = \frac{\omega}{4\pi c} 
\left( {\rm i} \frac{\partial}{\partial x} A_x - q_y A_y - q_z A_z \right) ,
\end{equation}
where the abbreviated notation $A_j\equiv A_j(\omega,{\bf q};x)$ is used.
The vector components of the electric current density, defined by
$4\pi {\bf j}(t,{\bf r}) = c\ {\rm curl}\, {\bf B} - \partial_t {\bf E}$,
are expressible as
\begin{eqnarray}
j_x(\omega,{\bf q};x) & = & \frac{c}{4\pi} \left[
\left( q^2 - \frac{\omega^2}{c^2} \right) A_x
- {\rm i} \frac{\partial}{\partial x} \left( q_y A_y + q_z A_z \right) 
\right] , \label{2.11} \\
j_y(\omega,{\bf q};x) & = & \frac{c}{4\pi} \left[
- {\rm i} q_y \frac{\partial}{\partial x} A_x 
+ \left( q_z^2 - \frac{\omega^2}{c^2} - \frac{\partial^2}{\partial x^2}
\right) A_y - q_y q_z A_z \right]  , \phantom{eee} \label{2.12} \\
j_z(\omega,{\bf q};x) & = & \frac{c}{4\pi} \left[
- {\rm i} q_z \frac{\partial}{\partial x} A_x - q_y q_z A_y
+ \left( q_y^2 - \frac{\omega^2}{c^2} - \frac{\partial^2}{\partial x^2}
\right) A_z \right]  . \phantom{eee} \label{2.13}
\end{eqnarray}

\renewcommand{\theequation}{3.\arabic{equation}}
\setcounter{equation}{0}

\section{Dipole sum rules} \label{Sect.3}
In the present and subsequent sections, we treat the symmetrized
charge-charge density correlation function
$\langle\rho(t,{\bf r}) \rho(0,{\bf r}')\rangle^s$, 
where we set $t'=0$ for simplicity.
This function fulfills the obvious neutrality condition
\begin{equation} \label{3.1}
\int {\rm d}{\bf r} \langle\rho(t,{\bf r}) \rho(0,{\bf r}')\rangle^s
= \int {\rm d}{\bf r}' \langle\rho(t,{\bf r}) \rho(0,{\bf r}')\rangle^s = 0.
\end{equation}

Let us consider a (partial) dipole moment carried by the charge-charge density
correlation function $\langle\rho(t,{\bf r}) \rho(0,{\bf r}')\rangle^s$,
with the point ${\bf r}$ being constrained to the region $\Lambda_1$:
\begin{equation} \label{3.2}
D^{(1)}(t) = \int_{-\infty}^{\infty}{\rm d}x' \int_0^{\infty} {\rm d}x
\int {\rm d}{\bf R}\, x \langle\rho(t,{\bf r}) \rho(0,{\bf r}')\rangle^s .
\end{equation}
Note that, due to the translational invariance of the correlation function
in the space perpendicular to the $x$ axis, the integration 
$\int {\rm d}{\bf R}$ can be equivalently rewritten as 
$\int {\rm d}{\bf R}'$ or $\int {\rm d}({\bf R}-{\bf R}')$.
Interchanging naively the order of integrations over $x'$ and $x$ in 
(\ref{3.2}), rewriting $\int {\rm d}{\bf R}$ as $\int {\rm d}{\bf R}'$  
and then applying the neutrality condition (\ref{3.1}), the quantity 
$D^{(1)}(t)$ seems to vanish.
This is not true.
As positive $x$ and $x'$ become large, 
$\langle\rho(t,{\bf r}) \rho(0,{\bf r}')\rangle^s$ tends to the bulk function 
$S^{(1)}_b(t,{\bf r}-{\bf r}')$ corresponding to the medium 1.
This correlation function is not small when the points ${\bf r}$ and 
${\bf r}'$ are close to one another.
Consequently, the function in (\ref{3.2}) is not absolutely integrable 
which prevents from permuting the integrations over $x'$ and $x$.
Subtracting and adding the bulk correlation function in (\ref{3.2}) leads to
\begin{eqnarray} 
D^{(1)}(t) & = & \int_{-\infty}^{\infty}{\rm d}x' \int_0^{\infty} {\rm d}x
\int {\rm d}{\bf R}\, x 
\left[ \langle\rho(t,{\bf r}) \rho(0,{\bf r}')\rangle^s -
S^{(1)}_b(t,{\bf r}-{\bf r}') \right] \nonumber \\
& & + \int_{-\infty}^{\infty}{\rm d}x' \int_0^{\infty} {\rm d}x
\int {\rm d}{\bf R}\, x S^{(1)}_b(t,{\bf r}-{\bf r}') . \label{3.3}
\end{eqnarray} 
We assume that the convergence of the charge-charge density correlation 
function to the bulk function occurs on a microscopic scale, so that
\begin{equation} \label{3.4}
\int_{-\infty}^{\infty}{\rm d}x' \int_0^{\infty} {\rm d}x  
\int {\rm d}{\bf R}\, x \vert \langle\rho(t,{\bf r}) \rho(0,{\bf r}')\rangle^s 
- S^{(1)}_b(t,{\bf r}-{\bf r}') \vert < \infty ;
\end{equation}
negative values of $x'$ do not represent any complication for $x>0$ since
the charge-charge density correlation function is expected to be short ranged
along the normal to the interface.
Under condition (\ref{3.4}), we can permute the $x'$ and $x$ integrals
in the first term on the r.h.s. of (\ref{3.3}); regarding the neutrality
sum rule (\ref{3.1}), this term becomes equal to 0.
Using the translation plus rotation invariance of the bulk correlation
function $S^{(1)}_b(t,{\bf r}-{\bf r}')$ and the method of integration 
by parts, the second term on the r.h.s. of (\ref{3.3}) 
can be easily reexpressed as follows
\begin{eqnarray}
\int_{-\infty}^{\infty}{\rm d}x' \int_0^{\infty} {\rm d}x
\int {\rm d}{\bf R}\, x S^{(1)}_b(t,{\bf r}-{\bf r}') 
& = & \frac{1}{2} \int_{-\infty}^{\infty} {\rm d}x
\int {\rm d}{\bf R}\, x^2 S^{(1)}_b(t,{\bf r}) \nonumber \\
& = & \frac{1}{6} \int {\rm d}{\bf r}\, {\bf r}^2 S^{(1)}_b(t,{\bf r}) . 
\label{3.5}
\end{eqnarray}
The calculations of this paragraph can be summarized by the equality
\begin{equation} \label{3.6}
\int_{-\infty}^{\infty}{\rm d}x' \int_0^{\infty} {\rm d}x
\int {\rm d}{\bf R}\, x \langle\rho(t,{\bf r}) \rho(0,{\bf r}')\rangle^s
= \frac{1}{6} \int {\rm d}{\bf r}\, {\bf r}^2 S^{(1)}_b(t,{\bf r}) .
\end{equation}
Note that this dipole sum rule depends on the bulk characteristics
of the only one from the two media.

We can treat similarly the (partial) dipole moment carried by the 
charge-charge density correlation function 
$\langle\rho(t,{\bf r}) \rho(0,{\bf r}')\rangle^s$,
when the point ${\bf r}$ is constrained to the region $\Lambda_2$:
\begin{equation} \label{3.7}
D^{(2)}(t) = \int_{-\infty}^{\infty}{\rm d}x' \int_{-\infty}^0 {\rm d}x
\int {\rm d}{\bf R}\, x \langle\rho(t,{\bf r}) \rho(0,{\bf r}')\rangle^s .
\end{equation}
The procedure analogous to the one outlined in the previous paragraph 
results in
\begin{equation} \label{3.8}
\int_{-\infty}^{\infty}{\rm d}x' \int_{-\infty}^0 {\rm d}x
\int {\rm d}{\bf R}\, x \langle\rho(t,{\bf r}) \rho(0,{\bf r}')\rangle^s
= - \frac{1}{6} \int {\rm d}{\bf r}\, {\bf r}^2 S^{(2)}_b(t,{\bf r}) ,
\end{equation}
where $S^{(2)}_b(t,{\bf r})$ is the bulk charge-charge density correlation 
function corresponding to the medium 2.
Combining relations (\ref{3.6}) and (\ref{3.8}), the total dipole moment reads
\begin{eqnarray}
\int_{-\infty}^{\infty}{\rm d}x' \int {\rm d}{\bf r}\, 
x \langle\rho(t,{\bf r}) \rho(0,{\bf r}')\rangle^s
& = & \int_{-\infty}^{\infty}{\rm d}x' \int {\rm d}{\bf r}\, (x-x') 
\langle\rho(t,{\bf r}) \rho(0,{\bf r}')\rangle^s \nonumber \\ & = &
\frac{1}{6} \int {\rm d}{\bf r}\, {\bf r}^2 
\left[ S^{(1)}_b(t,{\bf r}) - S^{(2)}_b(t,{\bf r}) \right] . \label{3.9}
\end{eqnarray}

We see that the dipole sum rules for an inhomogeneous configuration of
two semi-infinite media are expressible in terms of the 
second moments of the symmetrized charge-charge density correlation function 
in an infinite medium with the frequency-dependent dielectric function 
$\epsilon_1(\omega)$ or $\epsilon_2(\omega)$.
This subject was studied in Sect. 3 of the previous paper \cite{Samaj09}.
The final result for the second-moment condition, derived by using 
the Rytov fluctuational theory, reads
\begin{equation} \label{3.10}
\frac{\beta}{3} \int {\rm d}{\bf r}\, {\bf r}^2 S_b^{(\alpha)}(t,{\bf r})
= \int_{-\infty}^{\infty} \frac{{\rm d}\omega}{2\pi} 
\exp(-{\rm i}\omega t) \frac{g(\omega)}{\pi\omega}  
{\rm Im} \frac{1}{\epsilon_{\alpha}(\omega)} , 
\end{equation}
where the index $\alpha =1,2$ denotes the medium.
The introduced function
\begin{equation} \label{3.11}
g(\omega) \equiv \frac{\beta\hbar\omega}{2} 
\coth \left( \frac{\beta\hbar\omega}{2} \right)
\end{equation}
fulfills $g(\omega)\ge 1$, the equality $g(\omega)=1$ takes place
in the classical limit $\beta\hbar\omega\to 0$.
The integral over $\omega$ on the r.h.s. of (\ref{3.10}) is expressible 
in terms of elementary functions perhaps only for the (one-component)
jellium model of conductors, i.e. the system of identical particles
with the number density $n$, charge $e$ and mass $m$, immersed in a
neutralizing homogeneous background.
The dielectric function of the jellium is adequately described,
in the long-wavelength limit $q\to 0$, by the Drude formula with 
the dissipation constant taken as positive infinitesimal \cite{Jackson},
\begin{equation} \label{3.12}
\epsilon(\omega) = 1 - \frac{\omega_p^2}{\omega(\omega+{\rm i}\eta)} ,
\qquad \eta\to 0^+ ,
\end{equation}
where the plasma frequency $\omega_p$ is defined by 
$\omega_p^2 = 4\pi n e^2/m$.
Inserting the representation (\ref{3.12}) into (\ref{3.10}) and
using the Weierstrass theorem
\begin{equation} \label{3.13}
\lim_{\eta\to 0^+} \frac{1}{x\pm {\rm i}\eta} = 
{\cal P} \left( \frac{1}{x} \right) \mp {\rm i}\pi \delta(x)
\end{equation}
({\cal P} denotes the Cauchy principal value), we arrive at
\begin{equation} \label{3.14}
\frac{\beta}{3} \int {\rm d}{\bf r}\, {\bf r}^2 S_b(t,{\bf r})
= - \frac{1}{2\pi} g(\omega_p) \cos(\omega_p t) .
\end{equation}

In the static $t=0$ case, for all media, using path integration techniques 
and the general properties of dielectric functions in the complex
frequency upper half-plane, the integral over $\omega$ on the r.h.s. 
of (\ref{3.10}) can be formally expressed as \cite{Samaj09}
\begin{equation} \label{3.15}
\frac{\beta}{3} \int {\rm d}{\bf r}\, {\bf r}^2 S_b(0,{\bf r})
=  \frac{1}{2\pi} \left[ \frac{1}{\epsilon(0)} - 1 \right]
+ \frac{1}{\pi} \sum_{j=1}^{\infty} 
\left[ \frac{1}{\epsilon({\rm i}\xi_j)} - 1 \right] .
\end{equation}
Here, 
\begin{equation} \label{3.16}
\xi_j = \frac{2 \pi j}{\beta\hbar} \qquad (j=1,2,\ldots)
\end{equation}
are the (real) Matsubara frequencies.
For the general medium composed of species (electrons and ions) $\sigma$
with the number density $n_{\sigma}$, charge $e_{\sigma}$ and mass 
$m_{\sigma}$, the dielectric function fulfills the asymptotic relation
\cite{Jackson,LL}
\begin{equation} \label{3.17}
\epsilon(\omega) \mathop{\sim}_{\vert\omega\vert\to\infty}
= 1 - \frac{\omega_p^2}{\omega^2} , \qquad
\omega_p^2 = \sum_{\sigma} \frac{4\pi n_{\sigma} e_{\sigma}^2}{m_{\sigma}} .
\end{equation}
In the high-temperature (classical) limit $\beta\hbar\omega_p\to 0$,
each of the Matsubara frequencies $\{ \xi_j\}_{j=1}^{\infty}$ is much
larger than $\omega_p$, the corresponding terms in the sum on the r.h.s.
of (\ref{3.15}) vanish and so the formula (\ref{3.15}) represents
the split of the bulk second-moment condition onto its classical and purely 
quantum-mechanical parts.
We conclude that the dipole sum rules (\ref{3.6}) and (\ref{3.8}) take 
in the classical limit the following forms
\begin{eqnarray}
\beta \int_{-\infty}^{\infty}{\rm d}x' \int_0^{\infty} {\rm d}x
\int {\rm d}{\bf R}\, x \langle \rho({\bf r}) \rho({\bf r}')
\rangle^{\rm T}_{\rm cl} & = & \frac{1}{4\pi} 
\left( \frac{1}{\epsilon_1(0)} -1 \right) , \label{3.18} \\
\beta \int_{-\infty}^{\infty}{\rm d}x' \int_{-\infty}^0 {\rm d}x
\int {\rm d}{\bf R}\, x \langle\rho({\bf r}) \rho({\bf r}')
\rangle^{\rm T}_{\rm cl} & = & - \frac{1}{4\pi} 
\left( \frac{1}{\epsilon_2(0)} -1 \right) . \label{3.19}
\end{eqnarray}

In the above derivation of dipole sum rules, only the results of 
the bulk version of the Rytov fluctuational theory were needed at 
the final stage of the analysis. 
In what follows we shall show how Rytov's theory can be adopted from 
the beginning in its inhomogeneous version; the true value of this approach 
will be justified later.
Let the point ${\bf r}$ be in the region $\Lambda_1$, i.e. $x>0$,
the position of the point ${\bf r}'$ is arbitrary.
Using the formalism of Sect. 2 and the explicit results for 
the retarded Green function tensor in the Appendix, the Fourier transform 
of the charge-charge density correlation function is found to be
\begin{equation} \label{3.20}
\beta \langle \rho(x) \rho(x') \rangle^s_{\omega,{\bf q}} 
= - \frac{1}{2} \frac{g(\omega)}{\pi\omega} {\rm Im} \left[ 
\frac{1}{\epsilon_1(\omega)} \right]
\left( q^2 + \frac{\partial^2}{\partial x \partial x'} \right) \delta(x-x') .
\end{equation} 
Here, the delta function has to be understood in a macroscopic sense
(disregarding microscopic structure at small distances).
Taking in (\ref{3.20}) ${\bf q}={\bf 0}$ and performing the inverse Fourier 
transform in time, we obtain
\begin{eqnarray}
\beta \int {\rm d}{\bf R}\, \langle \rho(t,{\bf r}) \rho(0,{\bf r}') \rangle^s
& = & \frac{1}{2} \int_{-\infty}^{\infty} \frac{{\rm d}\omega}{2\pi} 
\exp(-{\rm i}\omega t) \frac{g(\omega)}{\pi\omega} 
{\rm Im} \left[ \frac{1}{\epsilon_1(\omega)} \right] \nonumber \\
& & \qquad \times \left[ - \frac{\partial^2}{\partial x \partial x'} 
\delta(x-x') \right] . \label{3.21}
\end{eqnarray}
Since the integration by parts implies
\begin{equation} \label{3.22}
\int_{-\infty}^{\infty} {\rm d}x' \int_0^{\infty} {\rm d}x\, x
\left[ - \frac{\partial^2}{\partial x \partial x'} \delta(x-x') \right] = 1 ,
\end{equation} 
we recover the dipole sum rule (\ref{3.6}) with the inserted bulk 
second-moment condition (\ref{3.10}).
The dipole sum rule (\ref{3.8}) can be verified analogously.

\renewcommand{\theequation}{4.\arabic{equation}}
\setcounter{equation}{0}

\section{Surface charge density correlations} \label{Sect.4}

\subsection{Classical limit}
We extend the classical result of \cite{Jancovici09b}, Sect. II, 
to the configuration in Fig. 1, i.e., the two half-spaces 
$\Lambda_1$ ($x>0$) and $\Lambda_2$ ($x<0$) filled with media characterized 
by static dielectric constants $\epsilon_1\equiv \epsilon_1(0)$ and 
$\epsilon_2\equiv \epsilon_2(0)$, respectively.
We recall that $\epsilon(0)\to {\rm i}\infty$ for conductors,
$\epsilon(0)=1$ for vacuum and $\epsilon(0)>1$ (finite) for dielectrics.
We shall consider the static two-point correlation functions with zero time
difference $t=t'$, so the time variables will be omitted in the notation. 

For an arbitrary configuration of two points ${\bf r}$ and ${\bf r}'$
in the media, we shall compute the correlation function 
$\langle\phi({\bf r})\phi({\bf r}')\rangle^{\rm T}$, where $\phi({\bf r})$ 
is the microscopic electric potential created by the media at point ${\bf r}$. 
It is related to the microscopic charge density $\rho({\bf r}'')$ by
\begin{equation} \label{4.1}
\phi({\bf r})=\int{\rm d}{\bf r}''\frac{\rho({\bf r}'')}{
\vert {\bf r}-{\bf r}''\vert} ,
\end{equation}
where the integral is over the whole space.
In particular, we shall first calculate microscopically the electric potential 
due to an infinitesimal charge $Q$ placed in one of the two media and 
then complete this calculation with the phenomenological electrostatics result 
(the method of images) for that potential.

Let us introduce a test infinitesimal pointlike charge $Q$ at point ${\bf r}$.
The microscopic formula for the {\em total} potential $\phi_{\rm tot}$ 
induced at point ${\bf r}'$ is
\begin{equation} \label{4.2}
\langle \phi_{\rm tot}({\bf r}') \rangle_Q = \frac{Q}{\vert {\bf r}'-{\bf r} 
\vert} + \langle \phi({\bf r}') \rangle_Q ,
\end{equation}
where $\langle\cdots\rangle_Q$ means an equilibrium average in the presence
of charge $Q$.
The additional Hamiltonian is $H=Q\phi({\bf r})$. 
We now use the linear response theory for $\langle \phi({\bf r}') \rangle_Q$
which says that
\begin{eqnarray}
\langle \phi({\bf r}') \rangle_Q & = & \langle \phi({\bf r}') \rangle  
- \beta \langle \phi({\bf r}')H \rangle^{\rm T} \nonumber \\
& = & \langle \phi({\bf r}') \rangle
-\beta Q \langle \phi({\bf r}')\phi({\bf r}) \rangle^{\rm T} , \label{4.3}
\end{eqnarray}
where $\langle\cdots\rangle = \langle\cdots\rangle_{Q=0}$ means the standard 
equilibrium average (i.e., in the absence of the test charge $Q$).
Combining (\ref{4.2}) and (\ref{4.3}) we arrive at
\begin{equation} \label{4.4}
\beta Q \langle \phi({\bf r}')\phi({\bf r}) \rangle^{\rm T} =
\frac{Q}{\vert {\bf r}'-{\bf r} \vert} -
\left[ \langle \phi_{\rm tot}({\bf r}') \rangle_Q - 
\langle \phi({\bf r}') \rangle \right] . 
\end{equation}

Now, let the point ${\bf r}$ be in region $\Lambda_1$.
According to phenomenological electrostatics \cite{Jackson}, the shift of 
the potential average due to $Q$ is in $\Lambda_1$ 
\begin{equation} \label{4.5}
\langle \phi_{\rm tot}({\bf r}') \rangle_Q - \langle \phi({\bf r}') \rangle 
= \frac{Q}{\epsilon_1 \vert {\bf r}'-{\bf r} \vert} +
\frac{Q'}{\epsilon_1\vert {\bf r}'-{\bf r}^{\ast}\vert} ,
\qquad Q' = \frac{\epsilon_1-\epsilon_2}{\epsilon_1+\epsilon_2} Q ,
\end{equation} 
where ${\bf r}^{\ast}=(-x,{\bf R})$ is the position of the image
charge $Q'$. 
We would like to emphasize that the relation (\ref{4.5}) is valid for 
{\em macroscopic} distances $\vert {\bf r}'-{\bf r} \vert$ which are much 
larger than the microscopic scale defined by the particle correlation length.
If the test charge is in region $\Lambda_2$ at ${\bf r}$, 
the average potential $\langle \phi({\bf r}') \rangle_Q$ in $\Lambda_2$ 
is given by (\ref{4.5}) with indices 1 and 2 interchanged. 
Finally, if the test charge is in region $\Lambda_1$ at ${\bf r}$, 
the average potential in region $\Lambda_2$ is given by
\begin{equation} \label{4.6}
\langle \phi_{\rm tot}({\bf r}') \rangle_Q - \langle \phi({\bf r}') \rangle 
= \frac{Q''}{\epsilon_2\vert {\bf r}'-{\bf r} \vert} , \qquad
Q'' = \frac{2\epsilon_2}{\epsilon_1+\epsilon_2} Q .
\end{equation}
A similar relation is also valid if the test charge is in $\Lambda_2$ for the 
average potential in region $\Lambda_1$.

Using (\ref{4.4}) and (\ref{4.5}), we obtain
\begin{equation} \label{4.7}
\beta \langle \phi({\bf r})\phi({\bf r}') \rangle^{\rm T} =
\left(1-\frac{1}{\epsilon_1}\right) \frac{1}{\vert {\bf r}'-{\bf r} \vert}
+\frac{\epsilon_2-\epsilon_1}{\epsilon_2+\epsilon_1} 
\frac{1}{\epsilon_1\vert {\bf r}'-{\bf r}^{\ast} \vert} \quad
\mbox{if $x,x'>0$.}
\end{equation} 
If $x,x'<0$, 1 and 2 should be interchanged. 
Similarly, (\ref{4.4}) and (\ref{4.6}) imply
\begin{equation} \label{4.8} 
\beta \langle \phi({\bf r})\phi({\bf r}') \rangle^{\rm T} = 
\left( 1 - \frac{2}{\epsilon_1+\epsilon_2} \right)
\frac{1}{\vert {\bf r}'-{\bf r}\vert } \quad
\mbox{if $x>0,x'<0$ or $x<0,x'>0$.}
\end{equation}

The surface charge $\sigma({\bf R})$ on the plane $x=0$ at the point 
$(0,{\bf R})$ is related to the discontinuity of the normal $x$-component
of the microscopic electric field ${\bf E}$ on the interface:
\begin{equation} \label{4.9}
4\pi\sigma({\bf R}) = E_x^+({\bf R}) - E_x^-({\bf R}),
\end{equation}
where the superscript $+$ $(-)$ means approaching the surface through 
the limit $x\rightarrow 0^+$ $(x\rightarrow 0^-)$. 
The surface charge correlation thus is 
\begin{eqnarray} 
\langle \sigma({\bf R})\sigma({\bf R}') \rangle^{\rm T} & = &
\frac{1}{(4\pi)^2}
\langle E_x^+({\bf R}) E_x^+({\bf R}') + E_x^-({\bf R}) E_x^-({\bf R}')
\nonumber \\ & & \qquad
-2 E_x^+({\bf R}) E_x^-({\bf R}') \rangle^{\rm T} . \label{4.10}   
\end{eqnarray}
The electric field is related to the potential by 
${\bf E}({\bf r})=-{\bf \nabla}\phi({\bf r})$, so that
\begin{equation} \label{4.11}
\langle E_x({\bf r}) E_x({\bf r}') \rangle^{\rm T} =
\frac{\partial^2}{\partial x\partial x'}
\langle \phi({\bf r}) \phi({\bf r}') \rangle^{\rm T} .
\end{equation}
Using (\ref{4.7}), we obtain
\begin{eqnarray}  
\beta \langle E_x^+({\bf R}) E_x^+({\bf R}') \rangle^{\rm T} & = &
\frac{\partial^2}{\partial x \partial x'}
\left[ \left( 1-\frac{1}{\epsilon_1}\right)\frac{1}{
\vert {\bf r}'-{\bf r}\vert } \right. \nonumber \\ & & \left. \qquad
+ \frac{\epsilon_2-\epsilon_1}{\epsilon_2+\epsilon_1}
\frac{1}{\epsilon_1\vert {\bf r}'-{\bf r}^{\ast}\vert } 
\right] \Bigg\vert_{x=x'=0} . \label{4.12}
\end{eqnarray}
Since
$$\frac{\partial^2}{\partial x \partial x'} 
\frac{1}{\vert {\bf r}'- {\bf r} \vert} \Big\vert_{x=x'=0} = 
\frac{1}{\vert {\bf R}-{\bf R}'\vert^3} , \quad 
\frac{\partial^2}{\partial x \partial x'} 
\frac{1}{\vert {\bf r}'- {\bf r}^{\ast} \vert} \Big\vert_{x=x'=0} = 
\frac{-1}{\vert {\bf R}-{\bf R}'\vert^3} , $$
we find
\begin{equation}  \label{4.13}
\beta \langle E_x^+({\bf R}) E_x^+({\bf R}') \rangle^{\rm T} =
\left(1-\frac{2}{\epsilon_1}+\frac{2}{\epsilon_1
+\epsilon_2}\right)\frac{1}{\vert {\bf R}-{\bf R}'\vert^3} .
\end{equation}
$\beta \langle E_x^-({\bf R})E_x^-({\bf R}')\rangle^{\rm T}$ is obtained from 
(\ref{4.13}) by interchanging 1 and 2. 
Finally,
\begin{equation}  \label{4.14}
\beta \langle E_x^+({\bf R})E_x^-({\bf R}')\rangle^{\rm T} =
\left(1-\frac{2}{\epsilon_1+\epsilon_2}\right)
\frac{1}{\vert {\bf R}-{\bf R}'\vert^3}.
\end{equation}
Using these relations in (\ref{4.10}) gives the classical result
\begin{equation}  \label{4.15}
\beta \langle \sigma({\bf R})\sigma({\bf R}') \rangle^{\rm T}_{\rm cl} 
= \frac{h_{\rm cl}(0)}{\vert {\bf R}-{\bf R}'\vert^3} , \qquad
h_{\rm cl}(0) = -\frac{1}{8\pi^2}\left( \frac{1}{\epsilon_1} +
\frac{1}{\epsilon_2}-\frac{4}{\epsilon_1+\epsilon_2} \right) ,
\end{equation}
where the argument of the prefactor $h_{\rm cl}(t-t')$ equals to 0
for the considered static case $t=t'$.
We recall that this classical static result is valid for asymptotic 
distances $\vert {\bf R}-{\bf R}' \vert$ much larger than any microscopic 
length scale.
If in $\Lambda_2$ there is vacuum ($\epsilon_2=1$), one retrieves 
equation (20) in \cite{Jancovici09b}. 
If furthermore in $\Lambda_1$ there is a conductor ($\epsilon_1=\infty$), 
one retrieves the old result of \cite{Jancovici82b}.     

The surface charge density $\sigma$ has to be understood as being
the volume charge density $\rho$ integrated along the $x$ axis on 
some microscopic distance within the interface region.
From this point of view, the formula (\ref{4.15}) implies a sum
rule for the volume charge-charge density correlation function.
In particular, if one assumes an asymptotic behavior of the form
\begin{equation}  \label{4.16}
\beta \langle \rho({\bf r})\rho({\bf r}') \rangle^{\rm T}_{\rm cl} =
\frac{h(x,x')}{\vert {\bf R}-{\bf R}'\vert^3} , \qquad
\vert {\bf R}-{\bf R}'\vert \to \infty ,
\end{equation}
$h(x,x')$ obeys, in the classical limit, the sum rule
\begin{equation} \label{4.17}
\int_{-\infty}^{\infty} {\rm d}x' \int_{-\infty}^{\infty} {\rm d}x\,
h(x,x') = h_{\rm cl}(0) .
\end{equation}

The above formalism can be extended straightforwardly to other geometries of 
the interface between media, e.g. cylindrically or spherically layered media, 
or to planarly multi-layered media.
The only modification consists in the application of the corresponding
variant of the method of images.

It is instructive to compare the present classical result (\ref{4.15}), 
valid for two fluctuating media, with the previous result \cite{Jancovici82b}
valid for a fluctuating medium in contact with the inert wall which 
``produces'' the images, but does not fluctuate.
For the special case of a Coulomb conductor $(\epsilon_1\to\infty)$
in contact with the fluctuating wall of the static dielectric constant
$\epsilon_2\equiv \epsilon_W$, the prefactor $h_{\rm cl}(0)$ in 
the formula (\ref{4.15}) takes the form
\begin{equation} \label{4.18}
\mbox{fluctuating wall:} \qquad
h_{\rm cl}(0) = - \frac{1}{8\pi^2} \frac{1}{\epsilon_W} .
\end{equation}
On the other hand, for a Coulomb conductor in contact with the inert wall
of the static dielectric constant $\epsilon_W$, the prefactor $h_{\rm cl}(0)$
was found to be \cite{Jancovici82b}
\begin{equation} \label{4.19}
\mbox{inert wall:} \qquad
h_{\rm cl}(0) = - \frac{1}{8\pi^2} \epsilon_W .
\end{equation}
We see that the two results (\ref{4.18}) and (\ref{4.19}) coincide with one
anther only for the vacuum (plain hard) wall; in vacuum, there are no charges, 
so that the description by fluctuating and inert walls should lead to 
the same result.
Increasing $\epsilon_W$ beyond 1, our formula (\ref{4.18}) predicts 
the suppression of the surface charge fluctuations while (\ref{4.19}) 
predicts their enhancement. 

\subsection{Classical Surface Charge Correlations and Dipole Moment}
In the classical limit, there exists a direct relation between the dipole
moments (\ref{3.18}), (\ref{3.19}) and the asymptotic behavior of
the surface charge density correlations (\ref{4.15}).
The aim of the present part is to derive this relation. 

Let us consider the potential-potential correlation function, given by
(\ref{4.7}) or (\ref{4.8}), when the point ${\bf r}$ is localized
at the interface, say ${\bf r}={\bf 0}$, the position of the point
${\bf r}'$ is arbitrary:
\begin{equation} \label{4.20}
\beta \langle \phi({\bf 0}) \phi({\bf r}') \rangle^{\rm T}
= \left( 1 - \frac{2}{\epsilon_1+\epsilon_2} \right)
\frac{1}{\vert {\bf r}'\vert} .
\end{equation}
Applying the Laplacian to both sides of this equation and using the
Poisson equation $\Delta_{{\bf r}'} \phi({\bf r}') = - 4\pi\rho({\bf r}')$, 
we get
\begin{equation} \label{4.21}
\beta \langle \phi({\bf 0}) \rho({\bf r}') \rangle^{\rm T}
= \left( 1 - \frac{2}{\epsilon_1+\epsilon_2} \right) \delta({\bf r}') .
\end{equation}
With regard to the definition of the microscopic potential (\ref{4.1}),
using in (\ref{4.21}) the (partial, two-dimensional) Fourier transform of 
the Coulomb potential
\begin{equation} \label{4.22}
\frac{1}{\vert {\bf r}\vert} = \int \frac{{\rm d}^2q}{(2\pi)^2}
{\rm e}^{{\rm i}{\bf q}\cdot {\bf R}} 
\frac{2\pi}{q} {\rm e}^{-q\vert x\vert} , \qquad q=\vert {\bf q}\vert ,
\end{equation}
and the convolution theorem, we get
\begin{equation} \label{4.23}
\beta \int_{-\infty}^{\infty} {\rm d}x\, 
\langle \rho(x) \rho(x') \rangle_{{\bf q}}^{\rm T} {\rm e}^{-q\vert x\vert}
= \frac{q}{2\pi} \left( 1 - \frac{2}{\epsilon_1+\epsilon_2} \right) \delta(x').
\end{equation}
This equation is valid for large distances $\vert {\bf R}-{\bf R}'\vert$
or, equivalently, small $q$.
Performing the small-$q$ expansion in (\ref{4.23}) and then integrating
over all $x'\in (-\infty,\infty)$, we arrive at  
\begin{eqnarray} 
\beta \int_{-\infty}^{\infty} {\rm d}x' \int_{-\infty}^{\infty} {\rm d}x\, 
\langle \rho(x) \rho(x') \rangle_{{\bf q}}^{\rm T} & = & 
\frac{q}{2\pi} \left( 1 - \frac{2}{\epsilon_1+\epsilon_2} \right) \nonumber \\
& & + q \beta \int_{-\infty}^{\infty} {\rm d}x' \int_{-\infty}^{\infty} 
{\rm d}x\, \vert x\vert \langle \rho(x) \rho(x') 
\rangle_{{\bf q}={\bf 0}}^{\rm T} .  \nonumber \\ \label{4.24}
\end{eqnarray}
This is the wanted relation.
Inserting here the relations for the dipole moments (\ref{3.18}) and 
(\ref{3.19}), we end up with
\begin{equation} \label{4.25}
\beta \int_{-\infty}^{\infty} {\rm d}x' \int_{-\infty}^{\infty} {\rm d}x\, 
\langle \rho(x) \rho(x') \rangle_{{\bf q}}^{\rm T} = \frac{q}{4\pi} 
\left( \frac{1}{\epsilon_1} + \frac{1}{\epsilon_2} 
- \frac{4}{\epsilon_1+\epsilon_2} \right) .
\end{equation}
Since, in the sense of distributions, the two-dimensional Fourier
transform of $1/R^3$ is $-2\pi q$, the result (\ref{4.25}) is equivalent
to the previous one described by equations (\ref{4.15})--(\ref{4.17}). 

\subsection{A Short Recapitulation of the Quantum Case}
The long-range decay of the quantum surface charge density correlation 
functions, in both retarded and nonretarded regimes, was the subject of
Refs. \cite{Samaj08,Jancovici09a,Jancovici09b}.
By using the Rytov formalism for a plane between two media, the two-point
electric field correlations were derived for any point positions in medium
1 and 2 and the discontinuity of the electric field across the interface
was related to the surface charge density.
The consequent integrals over the frequency were treated using path
integration techniques and the general properties of dielectric
functions in the complex frequency upper half-plane.

In the static $t=t'$ case, the final result for the Fourier transform of
the quantum surface charge density correlation function reads
\begin{equation} \label{4.26}  
\beta \langle \sigma \sigma \rangle_{{\bf q}} = \frac{q}{4\pi} 
\left( \frac{1}{\epsilon_1} + \frac{1}{\epsilon_2} 
- \frac{4}{\epsilon_1+\epsilon_2} \right) + F_{\rm qu}(0,q) ,
\end{equation}
where the explicit form of the (static) function $F_{\rm qu}(0,q)$ 
depends on the considered, retarded or nonretarded, regime.
In the retarded regime, we have
\begin{equation} \label{4.27}
F_{\rm qu}^{({\rm r})}(0,q) = \frac{q^2}{2\pi} \sum_{j=1}^{\infty}
\frac{1}{\kappa_1({\rm i}\xi_j) \epsilon_2({\rm i}\xi_j) +
\kappa_2({\rm i}\xi_j) \epsilon_1({\rm i}\xi_j)}
\frac{[\epsilon_1({\rm i}\xi_j) - \epsilon_2({\rm i}\xi_j)]^2}{
\epsilon_1({\rm i}\xi_j) \epsilon_2({\rm i}\xi_j)} ,
\end{equation}
where the Matsubara frequencies $\xi_j$ $(j=1,2,\ldots)$ are defined
in (\ref{3.16}) and the inverse lengths $\kappa_{\alpha}(\omega,q)$
for the regions $\alpha=1,2$ in (\ref{A.1}).
For the purely imaginary values of the frequencies $\omega={\rm i}\xi_j$,
the values of the dielectric functions $\epsilon_{1,2}({\rm i}\xi_j)$,
and consequently of the inverse lengths $\kappa_{1,2}({\rm i}\xi_j)$,
are real positive.
In the considered limit $q\to 0$, 
$\kappa_{1,2}({\rm i}\xi_j) = \xi_j \epsilon_{1,2}({\rm i}\xi_j)$.
Since $\xi_j\propto j$ and, according to (\ref{3.17}), 
$\epsilon_{1,2}({\rm i}\xi_j)-1 = O(1/j^2)$, the sum in (\ref{4.27})
converges. 
This means that the function $F_{\rm qu}^{({\rm r})}(0,q)$, being of 
the order $O(q^2)$, becomes negligible in comparison with the first term 
in (\ref{4.26}) in the limit $q\to 0$.
The prefactor associated with the asymptotic decay (\ref{4.15})
thus reads  
\begin{equation} \label{4.28}
h_{\rm qu}^{({\rm r})}(0) = -\frac{1}{8\pi^2}\left( \frac{1}{\epsilon_1} +
\frac{1}{\epsilon_2}-\frac{4}{\epsilon_1+\epsilon_2} \right) .
\end{equation}
This expression, which does not depend on the temperature and 
the Planck constant, coincides with the classical result (\ref{4.15}).
In other words, the consideration of retardation effects makes
the quantum mechanics equivalent to its classical limit.
The situation is fundamentally different in the nonretarded regime.
In the limit $c\to\infty$, it holds $\kappa_{\alpha}=q$. 
We thus get from the retarded representation (\ref{4.27}) that
\begin{equation} \label{4.29}    
F_{\rm qu}^{({\rm nr})}(0,q) = \frac{q}{2\pi} \sum_{j=1}^{\infty}
\left[ \frac{1}{\epsilon_1({\rm i}\xi_j)} + \frac{1}{\epsilon_2({\rm i}\xi_j)}
- \frac{4}{\epsilon_1({\rm i}\xi_j)+\epsilon_1({\rm i}\xi_j)} \right] .
\end{equation}
With regard to the asymptotic behavior
$\epsilon_{1,2}({\rm i}\xi_j)-1 = O(1/j^2)$, the sum in (\ref{4.29})
converges, the nonretarded function $F_{\rm qu}^{({\rm nr})}(0,q)$
is of the order $O(q)$ and therefore contributes to the surface charge 
density correlation (\ref{4.26}).
The prefactor $h_{\rm qu}^{({\rm nr})}(0)$ is a complicated function
of temperature.
It was shown in \cite{Samaj08} that for distances $\lambda\sim 1/q$
much smaller than $c/\omega_p$ the retardation effects are negligible
and so the nonretarded result (\ref{4.29}) takes place, while for 
$\lambda\gg c/\omega_p$ the retardation results describe adequately
the decay of the surface charge density correlations.   

In the retarded regime, the time difference between points has no effect
on the form of the asymptotic behavior (\ref{4.15}), i. e.
\begin{equation} \label{4.30}
h_{\rm qu}^{({\rm r})}(t) = -\frac{1}{8\pi^2}\left( \frac{1}{\epsilon_1} +
\frac{1}{\epsilon_2}-\frac{4}{\epsilon_1+\epsilon_2} \right) .
\end{equation} 

\renewcommand{\theequation}{5.\arabic{equation}}
\setcounter{equation}{0}

\section{Charge-current density correlations} \label{Sect.5}
In this section, we shall deal with the charge-current density correlation
functions $\langle \rho(t,{\bf r}) j_k(0,{\bf r}') \rangle^s$, 
where the component index $k$ equals $x\equiv x_1$, $y\equiv x_2$
or $z\equiv x_3$.
We recall that for the bulk medium with the dielectric function
$\epsilon(\omega)$, these correlations were shown to satisfy
the following sum rules \cite{Samaj09}
\begin{eqnarray}
\beta \int {\rm d}{\bf r}\, \langle \rho(t,{\bf r}) j_k(0,{\bf r}') 
\rangle^s_{\rm b} & = & 0 , \label{5.1} \\
\beta \int {\rm d}{\bf r}\, x_l \langle \rho(t,{\bf r}) j_k(0,{\bf r}') 
\rangle^s_{\rm b} & = & \delta_{kl} \int_{-\infty}^{\infty} 
\frac{{\rm d}\omega}{2\pi} \exp(-{\rm i}\omega t) \frac{g(\omega)}{2\pi{\rm i}}
{\rm Im} \frac{1}{\epsilon(\omega)} . \label{5.2}
\end{eqnarray}
The static $t=0$ version of the sum rule (\ref{5.2}) is trivial for
any medium:
Since $g(\omega) {\rm Im}\, \epsilon^{-1}(\omega)$ is an odd function
of $\omega$, the r.h.s. of (\ref{5.2}) vanishes.
In the special case of the jellium model with the dielectric function
(\ref{3.12}), the Weierstrass theorem (\ref{3.13}) permits us to
express explicitly the time-dependent sum rule (\ref{5.2}) as follows
\begin{equation} \label{5.3}
\beta \int {\rm d}{\bf r}\, x_l \langle \rho(t,{\bf r}) j_k(0,{\bf r}') 
\rangle^s_{\rm b} = \delta_{kl} \frac{g(\omega_p)\omega_p}{4\pi}
\sin(\omega_p t) .
\end{equation}

We now consider the inhomogeneous situation pictured in Fig 1.
Let the point ${\bf r}'$ be in the region $\Lambda_1$ $(x'>0)$,
the position of the point ${\bf r}$ is arbitrary.
We start with the inhomogeneous Rytov theory (see Sect. 2 and the Appendix),
whose results for the charge-current density correlation function in 
the Fourier space, up to terms linear in $q_y$ and $q_z$, 
can be summarized as follows
\begin{eqnarray}
\beta \langle \rho(x) j_x(x') \rangle^s_{\omega,{\bf q}}
& = & - \frac{g(\omega)}{2\pi{\rm i}} {\rm Im} \left[ 
\frac{1}{\epsilon_1(\omega)} \right] \frac{\partial}{\partial x} \delta(x-x') 
+ O(q_y^2,q_z^2,q_yq_z) , \label{5.4} \\
\beta \langle \rho(x) j_y(x') \rangle^s_{\omega,{\bf q}}
& = & - \frac{g(\omega)}{2\pi} {\rm Im} \left[ 
\frac{1}{\epsilon_1(\omega)} \right] q_y \delta(x-x') 
+ O(q_y^2,q_z^2,q_yq_z) , \label{5.5} \\
\beta \langle \rho(x) j_z(x') \rangle^s_{\omega,{\bf q}}
& = & - \frac{g(\omega)}{2\pi} {\rm Im} \left[ 
\frac{1}{\epsilon_1(\omega)} \right] q_z \delta(x-x') 
+ O(q_y^2,q_z^2,q_yq_z) . \label{5.6}
\end{eqnarray}

Taking ${\bf q}={\bf 0}$ in equations (\ref{5.4})--(\ref{5.6}) and 
regarding that
\begin{equation} \label{5.7}
\int_{-\infty}^{\infty} {\rm d}x \frac{\partial}{\partial x} \delta(x-x') 
= - \int_{-\infty}^{\infty} {\rm d}x \frac{\partial}{\partial x'} \delta(x-x') 
= 0 ,
\end{equation}
we obtain in the leading order
\begin{equation} \label{5.8}
\beta \int {\rm d}{\bf r}\, \langle \rho(t,{\bf r}) j_k(0,{\bf r}') \rangle^s
= 0  \qquad \mbox{for all $k=x,y,z$.}
\end{equation}
This is the analog of the bulk sum rule (\ref{5.1}) which holds also
for the point ${\bf r}'$ being situated inside the region $\Lambda_2$.

With respect to the equality (obtained with the aid of the integration by parts)
\begin{equation} \label{5.9}
\int_{-\infty}^{\infty} {\rm d}x\, x \frac{\partial}{\partial x} \delta(x-x') 
= - 1
\end{equation}
for (\ref{5.4}) and in the next order in $q$ of 
${\rm e}^{-{\rm i}{\bf q}\cdot({\bf R}-{\bf R}')}$ 
for the relations (\ref{5.5}), (\ref{5.6}), we find
\begin{equation} \label{5.10}
\beta \int {\rm d}{\bf r}\, x_l 
\langle \rho(t,{\bf r}) j_k(0,{\bf r}') \rangle^s = 
\delta_{kl} \int_{-\infty}^{\infty} 
\frac{{\rm d}\omega}{2\pi} \exp(-{\rm i}\omega t) \frac{g(\omega)}{2\pi{\rm i}}
{\rm Im} \frac{1}{\epsilon_1(\omega)} .
\end{equation}
This is the analog of the bulk sum rule (\ref{5.2}) for the point
${\bf r}'\in \Lambda_1$.
When ${\bf r}'\in \Lambda_2$, an analogous sum rule is obtained by 
substituting in (\ref{5.10}) $\epsilon_1(\omega)$ by $\epsilon_2(\omega)$.

There exist another sum rules for the inhomogeneous situation which
have no obvious counterpart in the bulk case.
These sum rules follow from the application of the equalities
\begin{eqnarray}
\int_{-\infty}^{\infty} {\rm d}x \int_0^{\infty} {\rm d}x'
\frac{\partial}{\partial x} \delta(x-x') =
- \int_{-\infty}^{\infty} {\rm d}x \int_0^{\infty} {\rm d}x'
\frac{\partial}{\partial x'} \delta(x-x') & = &  1 , \label{5.11} \\
\int_{-\infty}^{\infty} {\rm d}x \int_{-\infty}^0 {\rm d}x'
\frac{\partial}{\partial x} \delta(x-x') =
- \int_{-\infty}^{\infty} {\rm d}x \int_{-\infty}^0 {\rm d}x'
\frac{\partial}{\partial x'} \delta(x-x') & = &  - 1 , \phantom{aaaaaa}  
\label{5.12}
\end{eqnarray}
to the formula (\ref{5.4}) with $q=0$.
Namely, we have
\begin{equation} \label{5.13}
\beta \int {\rm d}{\bf R} \int_{-\infty}^{\infty} {\rm d}x 
\int_0^{\infty} {\rm d}x' \langle \rho(t,{\bf r}) j_x(0,{\bf r}') \rangle^s
= - \int_{-\infty}^{\infty} \frac{{\rm d}\omega}{2\pi}
{\rm e}^{-{\rm i}\omega t} \frac{g(\omega)}{2\pi{\rm i}}
{\rm Im} \frac{1}{\epsilon_1(\omega)} 
\end{equation} 
and, similarly,
\begin{equation} \label{5.14}
\beta \int {\rm d}{\bf R} \int_{-\infty}^{\infty} {\rm d}x 
\int_{-\infty}^0 {\rm d}x' \langle \rho(t,{\bf r}) j_x(0,{\bf r}') \rangle^s 
= \int_{-\infty}^{\infty} \frac{{\rm d}\omega}{2\pi}
{\rm e}^{-{\rm i}\omega t} \frac{g(\omega)}{2\pi{\rm i}} 
{\rm Im} \frac{1}{\epsilon_2(\omega)} . 
\end{equation}
These relations can be verified independently by using the method for 
the dipole sum rules developed in Sect. 3.
We first subtract from and add to the correlation function
$\langle \rho(t,{\bf r}) j_k(0,{\bf r}') \rangle^s$ on the l.h.s. 
of equations (\ref{5.13}) and (\ref{5.14}) its bulk counterparts,
corresponding to medium 1 if $x'>0$ and to medium 2 if $x'<0$.
As before, assuming that
\begin{equation} \label{5.15}
\int {\rm d}{\bf R} \int_{-\infty}^{\infty}{\rm d}x \int_0^{\infty} {\rm d}x'\,
\vert \langle\rho(t,{\bf r}) j_x(0,{\bf r}')\rangle^s 
- \langle\rho(t,{\bf r}) j_x(0,{\bf r}')\rangle^{s(1)}_{\rm b} \vert < \infty
\end{equation}
and, similarly,
\begin{equation} \label{5.16}
\int {\rm d}{\bf R} \int_{-\infty}^{\infty}{\rm d}x 
\int_{-\infty}^0 {\rm d}x'\, 
\vert \langle\rho(t,{\bf r}) j_x(0,{\bf r}')\rangle^s 
- \langle\rho(t,{\bf r}) j_x(0,{\bf r}')\rangle^{s(2)}_{\rm b} \vert < \infty ,
\end{equation}
the permutation of the $x$ and $x'$ integrations nullifies the contribution 
of the correlation function minus its bulk counterpart due to the sum rule
(\ref{5.8}).
Using the translational plus rotational invariance of the bulk correlation 
function in the nonzero term, we obtain
\begin{equation} \label{5.17}
\int {\rm d}{\bf R} \int_{-\infty}^{\infty} {\rm d}x \int_0^{\infty} 
{\rm d}x' \langle \rho(t,{\bf r}) j_x(0,{\bf r}') \rangle^{s(1)}_{\rm b}
= - \int {\rm d}{\bf r}\, x 
\langle \rho(t,{\bf r}) j_x(0,{\bf r}') \rangle^{s(1)}_{\rm b}
\end{equation}
and, similarly,
\begin{equation} \label{5.18}
\int {\rm d}{\bf R} \int_{-\infty}^{\infty} {\rm d}x \int_{-\infty}^0 
{\rm d}x' \langle \rho(t,{\bf r}) j_x(0,{\bf r}') \rangle^{s(2)}_{\rm b}
= \int {\rm d}{\bf r}\, x 
\langle \rho(t,{\bf r}) j_x(0,{\bf r}') \rangle^{s(2)}_{\rm b} .
\end{equation}
In view of these relations, the inhomogeneous sum rules 
(\ref{5.13}) and (\ref{5.14}) are in fact the consequences of 
the bulk sum rule (\ref{5.2}) for media 1 and 2, respectively. 

\renewcommand{\theequation}{6.\arabic{equation}}
\setcounter{equation}{0}

\section{Current-current density correlations} \label{Sect.6}
As concerns the current-current density correlations
$\langle j_k(t,{\bf r}) j_l(0,{\bf r}') \rangle^s$ $(k,l=x,y,z)$,
for the bulk medium with the dielectric function $\epsilon(\omega)$,
they satisfy the sum rule \cite{Samaj09}
\begin{equation} \label{6.1}
\beta \int {\rm d}{\bf r} 
\langle j_k(t,{\bf r}) j_l(0,{\bf r}') \rangle^s_{\rm b}
= - \delta_{kl} \int_{-\infty}^{\infty} \frac{{\rm d}\omega}{2\pi} 
\exp(-{\rm i}\omega t) \frac{g(\omega)\omega}{2\pi}
{\rm Im} \frac{1}{\epsilon(\omega)} .
\end{equation} 
In the case of the jellium model with the dielectric function (\ref{3.12}),
the Weierstrass theorem (\ref{3.13}) implies
\begin{equation} \label{6.2}
\beta \int {\rm d}{\bf r} 
\langle j_k(t,{\bf r}) j_l(0,{\bf r}') \rangle^s_{\rm b}
= \delta_{kl} \frac{g(\omega_p)\omega_p^2}{4\pi} \cos(\omega_p t) .
\end{equation}

In the static $t=0$ case, the formula (\ref{6.1}) can be formally
expressed as \cite{Samaj09}
\begin{equation} \label{6.3}
\beta \int {\rm d}{\bf r} 
\langle j_k({\bf r}) j_l({\bf r}') \rangle^s_{\rm b}
= \delta_{kl} \left\{ \frac{\omega_p^2}{4\pi} + \frac{1}{2\pi}
\sum_{j=1}^{\infty} \left[ \frac{\xi_j^2}{\epsilon({\rm i}\xi_j)}
- \xi_j^2 + \omega_p^2 \right] \right\} ,
\end{equation}
where $\xi_j$ are the Matsubara frequencies defined by (\ref{3.16}).
The first term on the r.h.s. of (\ref{6.3}) represents the classical
$\beta\hbar\omega_p\to 0$ limit, the second term is the purely 
quantum-mechanical contribution to the sum rule.

For the studied configuration in Fig. 1, the inhomogeneous version
of the Rytov method gives in the limit ${\bf q}\to {\bf 0}$,
for distinct current indices,
\begin{equation} \label{6.4}
\beta \langle j_k(x) j_l(x') \rangle^s_{\omega,{\bf q}={\bf 0}} = 0
\qquad \mbox{for $k\ne l$,}
\end{equation}
for any positions of points ${\bf r}$ and ${\bf r}'$ in media 1 and 2.
The relation (\ref{6.4}) is equivalent to
\begin{equation} \label{6.5}
\beta \int {\rm d}{\bf r} \langle j_k(t,{\bf r}) j_l(0,{\bf r}') \rangle^s
= 0 \qquad \mbox{for $k\ne l$,}
\end{equation}
where the position of the point ${\bf r}'$ in media 1 or 2 is irrelevant.
This is the generalization of the bulk sum rule (\ref{6.1}) for $k\ne l$.

Let the point ${\bf r}'$ be localized in the region $\Lambda_1$, i.e. $x'>0$,
the position of point ${\bf r}$ is arbitrary. 
For the $q\to 0$ limit of the diagonal correlation function of the $xx$ 
current components, the Rytov theory implies
\begin{equation} \label{6.6}
\beta \langle j_x(x) j_x(x') \rangle^s_{\omega,{\bf q}={\bf 0}} = 
- \frac{g(\omega)\omega}{2\pi} {\rm Im} \left[ 
\frac{1}{\epsilon_1(\omega)} \right] \delta(x-x') .
\end{equation}
Integrating over $x$, this equation gives
\begin{equation} \label{6.7}
\beta \int {\rm d}{\bf r} \langle j_x(t,{\bf r}) j_x(0,{\bf r}') \rangle^s
= - \int_{-\infty}^{\infty} \frac{{\rm d}\omega}{2\pi} 
\exp(-{\rm i}\omega t) \frac{g(\omega)\omega}{2\pi}
{\rm Im} \frac{1}{\epsilon_1(\omega)} , \quad {\bf r}'\in \Lambda_1 .
\end{equation} 
A similar expression can be derived when ${\bf r}'\in \Lambda_2$.

The inhomogeneous sum rules obtained up to now are quite trivial
generalizations of the corresponding bulk sum rule (\ref{6.1}).
This is no longer true for the diagonal correlation functions of the yy and
zz current components. 
For ${\bf r}'\in \Lambda_1$, the inhomogeneous Rytov method implies
\begin{eqnarray}
\beta \int {\rm d}{\bf r} \langle j_y({\bf r}) j_y({\bf r}') \rangle^s_{\omega}
& = & \beta \int {\rm d}{\bf r} \langle j_z({\bf r}) j_z({\bf r}') 
\rangle^s_{\omega} \nonumber \\ & = &
- \frac{g(\omega)\omega}{2\pi} {\rm Im} \left[ \frac{1}{\epsilon_1(\omega)}
+ f(\omega,x') \right] , \label{6.8}
\end{eqnarray}
where the additional position-dependent function $f(\omega,x')$, which
does not exist in the bulk case, reads
\begin{eqnarray}
f(\omega,x') & = & [1-\epsilon_1(\omega)] 
\frac{n_1(\omega)}{n_1(\omega)\epsilon_2(\omega)+n_2(\omega)\epsilon_1(\omega)} 
\frac{\epsilon_1(\omega)-\epsilon_2(\omega)}{\epsilon_1(\omega)}
\nonumber \\ & & \times
\exp\left( -\frac{\vert\omega\vert}{c} n_1(\omega) x'\right) \label{6.9}
\end{eqnarray}
with $n_{\alpha}(\omega)$ $(\alpha=1,2)$ defined by
\begin{equation} \label{6.10}
n_{\alpha}^2(\omega) = - \epsilon_{\alpha}(\omega) , \qquad
{\rm Re}\, n_{\alpha}(\omega) > 0 .
\end{equation}
The function $f(\omega,x')$ is equal to zero in three cases:
the homogeneous case $\epsilon_1(\omega) = \epsilon_2(\omega)$, 
the medium 1 is the trivial vacuum $\epsilon_1(\omega) = 1$ and far away
from the boundary $x'\to\infty$.
Since $\lim_{\omega\to\infty}f(\omega,x')=0$, the function $f(\omega,x')$
does not contribute to the classical limit of (\ref{6.8}), but it does 
contribute to quantum-mechanical corrections.
A similar expression can be derived when ${\bf r}'\in \Lambda_2$.

We conclude this section by noting that, according to the inhomogeneous
Rytov theory, the interface between two media breaks up the directional
invariance of the diagonal current-current correlations in the bulk.
While the sum rule for the xx correlations (\ref{6.7}) has the form
of the bulk one (\ref{6.1}), the yy and zz correlations (\ref{6.8}) 
exhibit an additional dependence on the distance from the interface.
We believe that this is not an artificial anomaly of the
applied method, but the true phenomenon occurring in the current-current
correlations functions. 

\renewcommand{\theequation}{7.\arabic{equation}}
\setcounter{equation}{0}

\section{Conclusion} \label{Sect.7}
In this paper, we applied the Rytov fluctuational electrodynamics
to the inhomogeneous geometry in Fig. 1 to derive 
a sequence of sum rules for the charge-charge, charge-current and
current-current density correlation functions. 
The validity of some of these sum rules was controlled independently 
by using methods developed previously in the context of the model of 
a fluctuating semi-infinite conductor in contact with an inert wall.

In the realistic model considered here, both semi-infinite media
in contact fluctuate.
Comparing the classical static results (\ref{4.18}) and (\ref{4.19})
for the fluctuating and inert walls, respectively, we see that
they coincide, as it should be, in the vacuum case $\epsilon_W=1$, 
but for $\epsilon_W>1$ these results are fundamentally different.

Some of the inhomogeneous sum rules represent a straightforward
generalization of their bulk counterparts.
This is not the case of the current-current density correlation functions;
the sum rules (\ref{6.7}) and (\ref{6.8}) indicate a breaking of 
the directional invariance of the diagonal current-current density 
correlations by the interface.

\begin{acknowledgements}
L. \v{S}. is grateful to LPT for very kind hospitality.
The support received from the European Science Foundation 
(``Methods of Integrable Systems, Geometry, Applied Mathematics''),
Grant VEGA No. 2/0113/2009 and CE-SAS QUTE is acknowledged. 
\end{acknowledgements}

\renewcommand{\theequation}{A.\arabic{equation}}
\setcounter{equation}{0}

\section*{Appendix} \label{Appendix}
In this Appendix, we present explicit forms of the retarded Green
function tensor elements $D_{jk}(\omega,{\bf q};x,x')$ for
the two semi-infinite media geometry pictured in Fig. 1. 
The half spaces $\Lambda_1$ $(x>0)$ and $\Lambda_2$ $(x<0)$ are
characterized, besides the dielectric functions $\epsilon_{\alpha}(\omega)$
$(\alpha=1,2)$, by the inverse lengths $\kappa_{\alpha}(\omega,q)$
$(\alpha=1,2)$ defined as follows
\begin{equation} \label{A.1} 
\kappa_{\alpha}^2(\omega,q) = q^2 - \frac{\omega^2}{c^2} 
\epsilon_{\alpha}(\omega), \qquad {\rm Re}\,\kappa_{\alpha}(\omega,q)>0. 
\end{equation}
Here, from the two possible solutions for each $\kappa_{\alpha}$ we choose 
the one with the positive real part in order to ensure the regularity
of tensor elements $D_{jk}(\omega,{\bf q};x,x')$ at asymptotically
large distances from the interface $x\to\pm\infty$. 
For simplification reasons, we shall omit in the notation the dependence of
functions on the frequency $\omega$ and the wave number $q$. 
\medskip

i) If the two points ${\bf r},{\bf r}'$ are localized in
the same half-space, say ${\bf r}',{\bf r} \in \Lambda_1$ (i.e. $x,x'>0$),
we introduce a pair of functions
\begin{eqnarray} 
u(x,x') & = & \frac{2\pi\hbar c^2}{\omega^2\epsilon_1\kappa_1}
\left[ {\rm e}^{-\kappa_1\vert x-x'\vert} + 
\frac{\epsilon_1\kappa_2-\epsilon_2\kappa_1}{\epsilon_1\kappa_2+
\epsilon_2\kappa_1} {\rm e}^{-\kappa_1(x+x')} \right] . \label{A.2} \\
v(x,x') & = & \frac{2\pi\hbar c^2}{\omega^2\epsilon_1\kappa_1}
\left[ {\rm e}^{-\kappa_1\vert x-x'\vert} -
\frac{\epsilon_1\kappa_2-\epsilon_2\kappa_1}{\epsilon_1\kappa_2+
\epsilon_2\kappa_1} {\rm e}^{-\kappa_1(x+x')} \right] . \label{A.3}
\end{eqnarray}
These functions satisfy the same type of the differential equation
\begin{equation} \label{A.4}
\left( \frac{\partial^2}{\partial x^2} - \kappa_1^2 \right) f 
= - \frac{4\pi\hbar c^2}{\omega^2\epsilon_1} \delta(x-x') ; \qquad
\mbox{$f = u(x,x')$ or $v(x,x')$,}
\end{equation}
and are related by
\begin{equation} \label{A.5}
\frac{\partial u(x,x')}{\partial x} = - \frac{\partial v(x,x')}{\partial x'} ,
\qquad
\frac{\partial u(x,x')}{\partial x'} = - \frac{\partial v(x,x')}{\partial x} .
\end{equation}
The third function we shall need is defined by
\begin{equation} \label{A.6}
w(x,x') = \frac{4\pi\hbar c^2}{\omega^2} \frac{\kappa_2-\kappa_1}{\kappa_1}
\frac{1}{\epsilon_1\kappa_2+\epsilon_2\kappa_1} {\rm e}^{-\kappa_1(x+x')} .
\end{equation}
In terms of the introduced functions, the elements of the retarded Green
function tensor are given by 
\begin{equation} \label{A.7}
D_{xx}(x,x') = \frac{\partial^2}{\partial x\partial x'} u(x,x')
- \frac{\omega^2}{c^2} \epsilon_1 v(x,x') ,
\end{equation}
\begin{equation} \label{A.8}
D_{xy}(x,x') = - {\rm i} q_y \frac{\partial}{\partial x} u(x,x') , \qquad
D_{yx}(x,x') = {\rm i} q_y \frac{\partial}{\partial x'} u(x,x') ,
\end{equation}
the remaining $xz$ and $zx$ components are given by the replacement rule
$D_{xz}(x,x') = D_{xy}(x,x') \{ q_y\leftrightarrow q_z \}, 
D_{zx}(x,x') = D_{yx}(x,x') \{q_y\leftrightarrow q_z \}$,
\begin{equation} \label{A.9}
D_{yy}(x,x') = \left[ q_y^2 - \frac{\omega^2}{c^2} \epsilon_1 \right]
u(x,x') + q_z^2 w(x,x') ,
\end{equation}
$D_{zz}(x,x') = D_{yy}(x,x') \{ q_y\leftrightarrow q_z \}$ and 
\begin{equation} \label{A.10}
D_{yz}(x,x') \equiv D_{zy}(x,x') = q_y q_z [ u(x,x') - w(x,x') ] .
\end{equation}
\medskip

ii) If the two points ${\bf r},{\bf r}'$ are localized in
the different half-spaces, say ${\bf r}'\in \Lambda_1$ and
${\bf r}\in \Lambda_2$ (i.e. $x'>0$ and $x<0$), we introduce the function
\begin{equation} \label{A.11}
s(x,x') = \frac{4\pi\hbar c^2}{\omega^2(\epsilon_1\kappa_2+
\epsilon_2\kappa_1)} {\rm e}^{\kappa_2 x - \kappa_1 x'} .
\end{equation}
In terms of this function, the elements of the retarded Green
function tensor are given by 
\begin{equation} \label{A.12}
D_{xx}(x,x') = - q^2 s(x,x') 
\end{equation}
\begin{equation} \label{A.13}
D_{xy}(x,x') = - {\rm i} q_y \kappa_1 s(x,x') , \qquad
D_{yx}(x,x') = - {\rm i} q_y \kappa_2 s(x,x') ,
\end{equation}
$D_{xz}(x,x') = D_{xy}(x,x') \{ q_y\leftrightarrow q_z \}, 
D_{zx}(x,x') = D_{yx}(x,x') \{ q_y\leftrightarrow q_z \}$,
\begin{equation} \label{A.14} 
D_{yy}(x,x') = [- q_z^2 + \kappa_1\kappa_2] s(x,x') ,
\end{equation}
$D_{zz}(x,x') = D_{yy}(x,x') \{ q_y\leftrightarrow q_z \}$ and 
\begin{equation} \label{A.15}
D_{yz}(x,x') = D_{zy}(x,x') = q_y q_z s(x,x') .
\end{equation}

\end{document}